\documentclass[prb,nopacs,twocolumn,preprintnumbers,amsmath,amssymb]{revtex4}

\usepackage{epsfig}
\usepackage{graphicx}
\usepackage{dcolumn}
\usepackage{color}
\usepackage{bm}
\newcommand{\vecr}{\vec{r}}

\newcommand{\veck}{\vec{k}}

\newcommand{\vf}{v_{\rm F}}

\newcommand{\half}{\mbox{\small $\frac{1}{2}$}}
\newcommand{\eexp}{\mbox{e}^}
\newcommand{\rr}{{\bf r}}

\begin{document}

\title{Gauge field induced by ripples in graphene.}
\author{F. Guinea{$^1$}, Baruch Horovitz{$^2$} and P. Le Doussal{$^3$} }

{\affiliation{\affiliation{$^1$} Instituto de Ciencia de Materiales
de Madrid. CSIC. Cantoblanco. E-28049 Madrid, Spain}
\affiliation{{$^2$} Department of Physics, Ben Gurion University,
Beer Sheva 84105 Israel}
 \affiliation{{$^3$} CNRS-Laboratoire de
Physique Th{\'e}orique de l'Ecole Normale Sup{\'e}rieure, 24 rue
Lhomond,75231 Cedex 05, Paris France.}

\begin{abstract}
We study the effects of quenched height fluctuations (ripples) in
graphene on the density of states (DOS). We show that at strong
ripple disorder a divergence in the DOS can lead to an ordered
ground state. We also discuss the formation of dislocations in
corrugated systems, buckling effects in suspended samples, and the
changes in the Landau levels due to the interplay between a real
magnetic field and the gauge potential induced by ripples.
\end{abstract}

\maketitle

\section{Introduction.}
The recent characterization of graphene sheets made up of a single layer of
carbon atoms\cite{Netal04,Netal05b} has caused great interest. Their unusual
electronic band structure, and the possibility of tuning the number of
electrons lead to a number of interesting features, both from a fundamental
perspective and because of its potential applications\cite{GN07,NGPNG07}.

The low energy electronic states of graphene are well described, in
the continuum limit, by two decoupled two dimensional Dirac
equations. The kinetic energy depends linearly on the lattice
momentum. The perturbations due to some types of disorder, like
topological lattice defects\cite{GGV92,GGV93b},
strains\cite{Metal06,MG06}, and curvature\cite{NK07} enter as an
effective gauge field. Curvature, strains and topological lattice
defects are expected to exist in graphene, as experiments show a
significant corrugation both in suspended samples\cite{Metal07}, in
samples deposited on a substrate\cite{Setal07b,CF07}, and also in
samples grown on metallic surfaces\cite{Vetal08}.

The statistical properties of the two dimensional Dirac equation
in a random gauge field have been extensively
studied\cite{LFSG94,CMW96,CCFGM97,HD02,K07}, in relation with the
Integer Quantum Hall effect. It has been shown that the density of
states develops a peak at zero energy when the disorder strength
exceeds a certain threshold. Furthermore, beyond a second
threshold, there is a transition to a glassy phase\cite{CD00}
where the local density of states is dominated by rare
regions.\cite{HD02}.

In the following, we will apply the analysis in\cite{HD02} to the
specific case of graphene, where there are two Dirac equations
coupled to the same random gauge field. The model will be detailed
in the next Section. We analyze in the following Section the
statistical properties of the gauge field. The main results for the
density of states are presented in Section IV.  Given a divergent
density of states at the energy of the Dirac point, we consider the
instabilities which may be induced by interactions. Alternative
approaches to the interplay between gauge fields and interactions
are given in\cite{W99,SGV05,HJV08}, although they did not consider
diverging densities of states. Sections VI analyze the related
problem of the structural changes which can be induced by the same
random strains which give rise to the gauge field, following the
analysis in\cite{CD98b}.  Section VII discusses a buckling
transition in suspended graphene. Section VIII estimates the effects
of ripples on density fluctuations in the quantum Hall regime and
compares with recent data \cite{Metal07b}. The main results of the
paper are summarized in Section VI.

\section{The model.}
We analyze the gauge field induced by the height fluctuations of a
graphene layer on a rough substrate. In such case one expects that the shape of the
graphene layer is determined by a competition between the interaction
of the layer with the rough substrate, which tends to impose a preferred
height, and the elastic properties of
the layer. A simple Hamiltonian which models these effects is:
\begin{eqnarray}
{\cal H} &= &{\cal H}_{subs} + {\cal H}_{lattice} + {\cal H}_{elec} \nonumber
\\
{\cal H}_{subs} &= &\frac{g}{2} \int d^2 \vecr \left[ h ( \vecr ) - h_0 (
  \vecr )  \right]^2 \nonumber \\
{\cal H}_{elastic} &= &\frac{\kappa}{2} \int d^2 \vecr \left[
    \nabla^2 h ( \vecr )
\right]^2  + \nonumber \\ &+ &\int d^2 \vecr \left\{  \frac{\lambda}{2} \left[
\sum_i u_{ii} ( \vecr ) \right]^2 +
\mu \sum_{ij} \left[ u_{ij} ( \vecr ) \right]^2 \right\} \nonumber \\
{\cal H}_{elec} &= &\vf \int d^2 \vecr  \bar{\Psi}_1 ( \vecr ) \left\{
    \sigma_x \left[ - i \partial_x - A_x ( \vecr ) \right] +
  \right. \nonumber \\ &+ &\left. \sigma_y \left[
      - i \partial_y - A_ y ( \vecr ) \right]\right\}
    \Psi_1 ( \vecr )  - \nonumber \\
&- &\vf \int d^2 \vecr  \bar{\Psi}_2 ( \vecr ) \left\{
    \sigma_x \left[ - i \partial_x + A_x ( \vecr ) \right] +
  \right. \nonumber \\ &+ &\left. \sigma_y \left[
      - i \partial_y + A_ y ( \vecr ) \right]\right\}
    \Psi_2 ( \vecr )
\label{hamil}
\end{eqnarray}
where $h ( \vecr)$ is the height of the graphene layer, $h_0( \vecr )$ is the preferred
height which can be assumed to follow closely the substrate height, $\Psi_1 ( \vecr )$ and
$\Psi_2 ( \vecr )$ are the two inequivalent Dirac (iso)-spinors
which can be defined in the graphene lattice, $u_{ij} ( \vecr )$
is the strain tensor associated to the deformation of the graphene layer, given by:
\begin{eqnarray}
u_{xx} &= &\frac{\partial u_x}{\partial x} + \half\left(
\frac{\partial
    h}{\partial x} \right)^2 \nonumber \\
u_{yy} &= &\frac{\partial u_y}{\partial y} + \half\left(
\frac{\partial
    h}{\partial y} \right)^2 \nonumber \\
u_{xy} &= &\frac{1}{2} \left( \frac{\partial u_x}{\partial y} +
    \frac{\partial u_y}{\partial x} \right) + \half\frac{\partial h}{\partial x}
    \frac{\partial h}{\partial y}
\label{strain}
\end{eqnarray}
The gauge vector acting on the electrons in eq.(\ref{hamil}) is
related to the strain tensor by\cite{SA02b,M07}:
\begin{eqnarray}
A_x ( \vecr ) &= &\frac{\beta}{a} \left[ u_{xx} ( \vecr ) - u_{yy} ( \vecr )
\right]
\nonumber \\
A_y ( \vecr ) &= &-2 \frac{\beta}{a} u_{xy} ( \vecr )
\label{gauge}
\end{eqnarray}
where $a \approx 1.4$\AA \, is the length of the bond between
neighboring carbon atoms, and $\beta=C \tilde \beta$ where $C$ is a constant of
order unity and $\tilde \beta = -\partial \log ( t ) /
\partial \log ( a ) \sim 2-3$ is a dimensionless parameter which
characterizes the coupling between the Dirac electrons and lattice deformations. Besides the symmetry arguments
in\cite{M07}, we also assume that the coupling between the electrons and lattice deformations is through the
modulation of the hopping between nearest neighbor $\pi$ orbitals, $t \approx 3$eV\cite{G81,HKSS88}. The rest of
the parameters which determine the hamiltonian in eq.(\ref{hamil}) are the electron Fermi velocity, $\vf = 3 t a
/ 2$, the bending rigidity, $\kappa \sim 1$eV, the in-plane elastic constants, $\lambda , \mu \sim 1$eV
\AA$^{-2}$. To model the interaction between the graphene layer and the substrate we use a simple quadratic
expansion around the height $h_0(\vec r)$ which minimizes the energy in the absence of elastic and electronic
energy, parameterized by a coupling $g$. The value of this parameter is less understood. Estimates based on the
analysis of the electrostatic potential between graphene and SiO$_2$\cite{Setal07b} suggest that $g \sim
10^{-2}-10^{-1}$meV \AA$^{-4}$. By comparing $g$ and $\kappa$, one finds that the pinning by the substrate
dominates for length scales greater than $l_{p} \sim (\kappa / g)^{1/4} \sim 10$\AA. The coupling $g$ being
strongly relevant, for $l \gg l_p$ it can be considered as effectively infinite and the graphene layer rigidly
pinned to the substrate, $h ( \vecr ) \approx h_0(\vecr)$ for $l \gg l_p$. Note that we assume here that effect
of direct pinning of the in plane modes by the substrate are small and can be neglected.

\section{Effective gauge field.}

Experiments\cite{Setal07,CF07} suggest that the height of the
graphene layer shows fluctuations of order $h \sim 10$\AA \, over
scales $l \sim 100$\AA. Similar fluctuations have been observed in
suspended graphene sheets\cite{Metal07}. We will assume that the
effects of the height fluctuations can be described statistically
over distances larger than $l_0 \sim 100$\AA. We will then relate the
correlations of the effective random gauge field to the (four point)
correlations of the (random) height profile, $h ( \vecr )$. The calculation
is valid whether this profile arises from interaction with
a static rough substrate (in which case for $l \gg l_p$ it directly relates
to substrate correlations) or from any other mechanism such as in
suspended graphene.

An estimate of the magnitude of the effective random gauge field
can be obtained by noting that the height change between
neighboring lattice points is $\sim a\nabla h$ hence the distance
change is $\sim a(\nabla h)^2$ and the modulation in $t$ is
$\delta t\sim \beta t(\nabla h)^2$. Hence the modulation in $A$ is
$\sim \delta t/v_F\sim \beta (\nabla h)^2/a$ which yields an
estimate for the variance of the random effective magnetic field
$B=[{\bm \nabla} \times {\bm A}]_z$:
\begin{eqnarray}
\langle B(q) B(q') \rangle && = C_B(q) (2 \pi)^2 \delta^2(q+q') \label{defB} \\
\pi \sigma &&= \lim_{q \to 0} q^{-2} C_B(q)
\nonumber\\&&\sim \left( \frac{\beta}{a} \right)^2 \int_{| \vecr |
  \le l_0}  d^2 \vecr \left(
  \frac{h}{l_0} \right)^4 \sim \frac{\beta^2 h^4}{a^2 l_0^2}
  \label{estimate}
\end{eqnarray}
where $h \sim 10$\AA \, is the typical scale of the height
fluctuations, as discussed earlier. Typical parameters allow for
$\sigma=O(1)$, within range of the transitions that we consider
below.

In order to perform a more detailed calculation of the effective
gauge field acting on the electrons, we first compute the in plane
displacement field, $\vec{u} ( \vecr )$ obtained by minimizing the
elastic energy for a given realization of $h ( \vecr )$, and then
we estimate the strain tensor $u_{ij} ( \vecr )$. We define:
\begin{equation}
f_{ij} ( \vecr ) = \frac{\beta}{a} \frac{\partial h}{\partial x_i}
  \frac{\partial h}{\partial x_j}
\end{equation}
In terms of these quantities, the procedure described above gives for the
effective magnetic field acting on the electrons:
\begin{eqnarray}
B ( \veck ) &= & i k_y \frac{( 3 k_x^2  -  k_y^2 ) ( \lambda + \mu
)}{( \lambda + 2 \mu )
  k^4} \times \nonumber \\
&\times & \left[ k_y^2 f_{xx} ( \veck ) + k_x^2 f_{yy} ( \veck ) - 2 k_x k_y
  f_{xy} ( \veck ) \right]
\label{bmag}
\end{eqnarray}
We assume
that the average properties of the height modulations
are described by translationally invariant correlation functions, in Fourier:
\begin{equation}
\left\langle f_{ij} ( \vec q ) f_{kl} ( \vec q ) \right\rangle =
{\cal F}_{ijkl}(q)  \label{corr}
\end{equation}
and are of short range character, i.e. with a finite limit for
$q l_0 \ll 1$:
\begin{equation}
{\cal F}_{ijkl}(q)|_{q \to 0} = f  \delta_{ij} \delta_{kl} +  f' \left(
\delta_{ik} \delta_{jl} + \delta_{il} \delta_{jl} \right)
\label{tensor}
\end{equation}
a tensor compatible with the hexagonal symmetry of the lattice
parameterized by two dimensionless constants $f$ and $f'$.
Using eqs.(\ref{bmag}), (\ref{corr}), and (\ref{tensor}), we find
for the correlations (\ref{defB}) of the effective magnetic field at small $q$:
\begin{equation}
C_B(q)
= q^2 \left(
  \frac{\lambda + \mu}{\lambda + 2 \mu} \right)^2 \sin^2 ( 3 \theta )
  ( f + 2 f' )  \label{gauge_B}
\end{equation}
where $q_x+i q_y = q e^{i \theta}$,
where the angle $\theta$ is measured from a given lattice
axis. The angular dependence of the correlation is consistent with
the lattice symmetry; as we show below, only its angular average
is relevant for the transitions.

In Eq. (\ref{corr}) we have assumed that the 2 point function of
$(\bm{\nabla}h)^2$ field has a finite $q=0$ limit. The
exact bound $\frac{\beta}{a} |\langle \partial_i h(\vec r) \partial_j h(\vec r') \rangle|
\leq |{\cal F}_{iijj}(\vec r-\vec r')|^{1/2}$ implies that the
roughness $h\sim r^{\zeta}$ of the graphene sheet (hence of the substrate if adsorbed)
can be at most $\zeta<1/2$ in the general case for Eq. (\ref{corr}) to hold. In a model
with Gaussian distributed $h$ the condition is $\zeta<1/4$ and higher roughness would result in long range (LR)
correlations in the disorder. Such LR correlations would
presumably arise when quenching thermal fluctuations of a freely
fluctuating membrane (which has $\zeta=0.59$ \cite{radz}) although a precise estimate
then requires taking into account non gaussian fluctuations, a non
trivial calculation. Here we restrict to SR disorder and
substrates such that Eq. (7) holds.

\section{Electronic density of states.}
We analyze the electronic density of states near the Dirac point,
$E=0$, using the techniques discussed in\cite{HD02}. The main
difference with the cases considered there is the existence of two
Dirac equations coupled to the same gauge field, with couplings of
equal absolute value but opposite sign, see ${\cal H}_{elec}$ in
eq.(\ref{hamil}).

The bosonized version of the problem also contains two fields,
which become two sets of coupled fields when the replica trick is
used to integrate over the disorder. Finally, we make the same
variational ansatz as in\cite{HD02}. The simplest observable is
the total density of states (DOS), which is self averaging and is
just twice the DOS of a single Dirac equation (single layer
problem as defined in \cite{HD02}) and behaves as:
\begin{equation}
\rho ( E ) \sim E^{2/z-1}
\label{dos}
\end{equation}
with:
\begin{equation}
z = \left\{ \begin{array}{lr} 2 - K +  \sigma K^2 &\sigma <
2/K^2 \\  K \left( \sqrt{8 \sigma} - 1 \right) &\sigma > 2/K^2  \end{array} \right.
      \label{exponent}
\end{equation}
where $K$ is a parameter which describes the kinetic energy of the
field in the bosonized version of the model, and, for the non
interacting case which corresponds to the hamiltonian in
eq.(\ref{hamil}) takes the value $K=1$. The parameter $\sigma$
determining the exponent in (\ref{exponent}) is found to be given
by the angle average of (\ref{gauge}):
\begin{equation}
 \sigma = \frac{1}{2 \pi} \left(
  \frac{\lambda + \mu}{\lambda + 2 \mu} \right)^2
  ( f + 2 f' )  \label{sigma}
\end{equation}
i.e. the strength of the random gauge field, consistent with the
order of magnitude estimate (\ref{estimate}).

The change in the dependence of the exponent $z$ on the strength
of the gauge field, $\sigma$, in eq.(\ref{exponent}) is associated
with a phase transition in the disordered bosonic model. For
$\sigma > \sigma_c = 2/K^2$ the local DOS (averaged over regions
of size up to $L\sim |E|^{-1/z}$) exhibits strong fluctuations and
non gaussian tails (i.e its disorder average being different from
its typical value) due to the dominance of rare regions. Note that
the divergence of the DOS at $E=0$ occurs at $\sigma = 1/K^2 <
\sigma_c$, i.e. before the freezing transition in the one layer
problem as disorder is increased.

\begin{figure}[t]
\begin{center}
\includegraphics[scale=0.25]{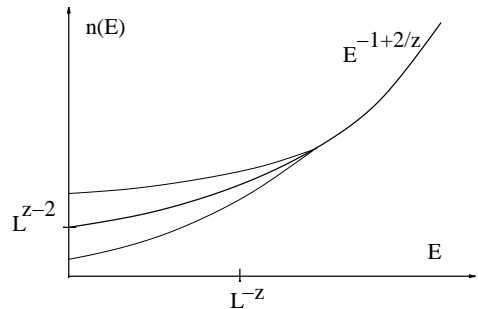}
\end{center}
\caption{Sketch of the DOS in a finite size $L$ region for $z<2$.
The thick line is a typical value, while the thin line represent the
size of fluctuations that are enhanced below the energy $L^{-z}$.
For $z>2$ the DOS increases at small $E$ and its typical value
saturates at $L^{-2+z}$. For $z>3$ (frozen regime) the fluctuations
become so strong that the average DOS grows as $L^{-2+{\bar z}}$
where ${\bar z}=1+\sigma >z$. Such finite size fluctuations should
be observable in tunneling experiments.}
\end{figure}

The effect on the DOS of an additional smooth random scalar
potential with variance $\delta$, corresponding to local
fluctuations of the chemical potential, induced by e.g. the
substrate, has been discussed in \cite{HD02}. It leads to:
\begin{equation}
 \rho(E) = E^{2/z-1} {\cal R}(E/\delta^{z/z'})  \label{w}
\end{equation}
where the exponent $z'$ is given by:
\begin{equation}
z' = \left\{ \begin{array}{lr} 2 - 2 K +  4 \sigma K^2 &\sigma <
\frac{1}{2
      K^2} \\  2 K \left( \sqrt{8 \sigma} - 1 \right) &\sigma >
      \frac{1}{2
      K^2} \end{array} \right.
      \label{exponent1}
\end{equation}
and exhibits a transition at $\sigma'_c=1/(2 K^2)=\sigma_c/4$.
This leads to a finite and non zero DOS at zero energy:
\begin{equation}
 \rho(E) \sim \delta^{(2-z)/z'}  \label{w1}
\end{equation}
a behavior which thus exhibits two distinct freezing transitions.
The divergence of the DOS at $\sigma=1$ (for $K=1$) is in between
these transitions.

Although we will not study this aspect in detail here, it is also
interesting to note that since the two Dirac equations describing
the two valleys (the two Fermi points) feel opposite random gauge
fields, mutual correlations of the local DOS in the two valleys as
measured by $\langle \rho_1(E,r) \rho_2(E,r) \rangle^c$ are
strong. They are found to exhibit a transition at a different
value of disorder $\sigma=1/(2 K^2)$ as can be seen by a study
analogous to the two layer model of Section IV B of \cite{HD02}

\begin{figure}[t]
\begin{center}
\includegraphics[scale=1]{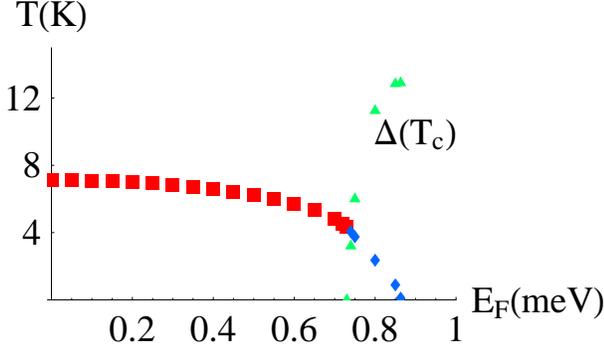}
\end{center}
\caption{(Color online). Critical temperature as function of
chemical potential. The parameters used are $W_0 = 200$meV, $l_0 =
10 a$, $\sigma = 1.4 (z=2.4)$, and $U = 1$eV. The value of $\sigma$
implies an average height fluctuation $h \approx 3.4$\AA. The blue
diamonds give the critical temperature when the transition is
discontinuous. The green triangles are the values of the gap
$\Delta$, in Kelvin, at the transition temperature, in the region
where $\Delta$ jumps discontinuously from zero to a finite value.}
\label{phase_diagram}
\end{figure}

\begin{figure}[t]
\begin{center}
\includegraphics[scale=0.4]{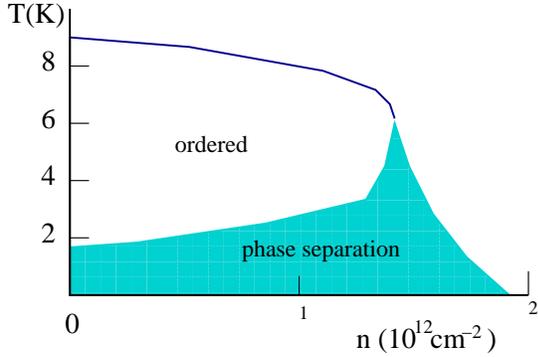}
\end{center}
\caption{(Color online). Approximate phase diagram, as function of
electron density and temperature, obtained with the same parameters
used in Fig.[\protect{\ref{phase_diagram}}]. The existence of a
first order transition leads to a region where electronic phase
separation is induced.} \label{phase_separation}
\end{figure}

\section{Interaction effects and electronic instabilities.}
For sufficiently large disorder, $\sigma > 1$, the density of
states, and, as a consequence, the electronic compressibility,
diverges at $E=0$. The electron-electron interaction, or the
interaction of the electrons with other degrees of freedom, will
lead to instabilities, which suppress the compressibility.

Within mean field theory, the effects of interactions on the
electronic band structure can be described as an external
potential which must be calculated self consistently. A simple
such potential which opens a gap at $E=0$ is the shift of the
energy on one sublattice of the honeycomb structure with respect
to the other. This shift can be associated to a spin or to a
charge density wave, or it can be induced by phonons\cite{FL07} or
by short range electron-electron
interactions\cite{ST92,GGV01,PAB04,H06}. In the continuum model
described here, it enters as a mass term:
\begin{eqnarray}
{\cal H}_{elec}^{tot} &= &{\cal H}_{elec} + {\cal H}_\Delta \nonumber \\
{\cal H}_\Delta &= &\Delta \sum_i \int d^2 \vecr \left[ \bar{\Psi}_i (
\vecr ) \sigma_z \Psi_i ( \vecr ) \right]
\end{eqnarray}
where ${\cal H}_{elec}$ is defined in eq.(\ref{hamil}).
The total electronic hamiltonian satisfies:
\begin{equation}
\left( {\cal H}_{elec} + {\cal H}_\Delta \right)^2 = \left( {\cal H}_{elec}
\right)^2 + \Delta^2 {\cal I}
\end{equation}
where ${\cal I}$ is the four dimensional unit matrix, independent
of spatial position, which acts on the space spanned by the four
component electronic (iso)-spinors. The eigenvalues of the
electronic hamiltonian satisfy ${\epsilon_n^{tot}}^2 =
\epsilon_n^2 + \Delta^2$, where $\epsilon_n$ is an eigenvalue of
${\cal H}_{elec}$. As a result, the density of states associated
to ${\cal H}_{elec}^{tot}$, $\rho_\Delta ( E )$,  satisfies:
\begin{equation}
\rho_\Delta ( E ) = \frac{E}{\sqrt{E^2 - \Delta^2}} \rho (
\sqrt{E^2 - \Delta^2})
\end{equation}
and, using the expression in eq.(\ref{dos}), we find:
\begin{equation}
\rho_\Delta ( E ) = \left\{ \begin{array}{lr} 0 &| E | < \Delta \\
    \frac{1}{l_0^2} \frac{E \left( \sqrt{E^2 - \Delta^2}
    \right)^{2/z-2}}{W_0^{2/z}} &W_0 > |E|>\Delta \end{array} \right.
\label{dos_D}
\end{equation}
where the energy scale $W_0 = \vf / l_0$ is inserted so that for
$E \gg W_0 \gg \Delta$ the value of $\rho_\Delta ( E )$ crosses
over into the density of states of the clean system, $\rho ( E )
\sim | E | / \vf^2$.

The self consistent value of $\Delta$ is determined by the
competition between the cost in energy associated to the formation
of the gap, and the decrease in electronic energy due to the
reduction in the density of states at the Fermi level. Near the
transition, $\Delta$ is small compared to the other energy scales
of the model, and the energy required to create the charge or spin
density wave can be expanded as function of $\Delta$. For
simplicity, we assume that the ordered phase is a spin density
wave induced by the on-site Hubbard repulsion, $U$ which breaks
the sublattice symmetry and produces a gap $\Delta=U S/2$ where
$S$ is the resulting polarization per site. The total energy is
the sum of the kinetic energy and the gain in interaction energy
obtained by inducing the polarization:
\begin{eqnarray}
E_{tot} &= &E_{elec} + E_{SDW} \nonumber \\
 E_{SDW} &= &  \frac{\Delta^2}{U} \nonumber \\
F_{el}&=&-4Ta^2\int_{\Delta}^{W_0}[\ln
(1+\eexp{(-E-E_F)/T})+\nonumber\\&&\ln
(1+\eexp{(E-E_F)/T})]\cdot\rho_{\Delta}(E)dE
\end{eqnarray}
%
%
where $4$ allows for spin and valley degeneracy and we allow for a
possibly non zero Fermi energy $E_F$. The induced gap is given by
minimizing the total free energy and with a change of integration
variable:
\begin{eqnarray}\label{sce}
1&=&2a^2U
\int_0^{W_0}\frac{\sinh(\sqrt{E^2+\Delta^2}/T)}{\cosh(E_F/T)+\cosh(\sqrt{E^2+\Delta^2}/T)}
\times\nonumber\\&& \frac{\rho(E)dE}{\sqrt{E^2+\Delta^2})}
\end{eqnarray}

We consider first the case $E_F = T = 0$ where $F_{el}\rightarrow
E_{el}=-4a^2\int_{\Delta}^{W_0}E\rho_{\Delta}(E)dE$. The integrand
 can be expanded for $E \gg
\Delta$, where it goes as $\Delta^2 E^{(2/z)-2}$. As a result, we
obtain a contribution to the electronic energy $\delta_1 E_{elec}
( \Delta ) \sim - ( a / l_0 )^2 \Delta^2 / W_0$. There is also a
contribution from the region $E \sim \Delta$. This term in
$E_{elec} ( \Delta )$ can be written as $\delta_2 E_{elec} (
\Delta ) \sim - ( a / l_0 )^2 \Delta^{(2/z)+1} W_0^{-2/z}$.

The relative strength of the two terms discussed above leads to
the existence of three regimes: i) $2/z-1 > 0$. The electronic
energy is determined by $\delta_1 E_{elec} ( \Delta )$. Both the
magnetic and electronic energy go as $\sim \Delta^2$, and, for $U
\ll \vf / a$ the minimum energy is at $\Delta=0$. ii) For $2/z-1 =
0$, we find $\delta_1 E_{elec} \sim -2( a / l_0 )^2\Delta^2 / W_0
\log ( W_0 / \Delta )$. The magnetic energy is greater by a
logarithmic factor, and there is an ordered phase, with $\Delta
\sim W_0 e^{- ( W_0 l_0^2 ) / ( 2U a^2 )}$. The problem becomes
equivalent to the Peierls analysis of the instability of a one
dimensional metal. iii) For $2/z-1 < 0$, the leading contribution
is $\delta_2 E_{elec}$. There is a magnetic phase with a gap:
\begin{equation}
\Delta_c \sim W_0  \left( \frac{a^2 U}{l_0^2 W_0}
\right)^{\frac{z}{z-2}}
\end{equation}

We now analyze the way in which the magnetic phase which always
exists for $2/z-1 <0$ is modified when $E_F , T \ne 0$. In
particular the order of the transition is determined by the sign
of the $a_4$ coefficient in the free energy expansion
$F=a_2\Delta^2+a_4\Delta^4$. Taking a $\partial_{\Delta^2}|_0$ on
the right hand side of Eq. (\ref{sce}) yields $a_4$, hence the
simultaneous conditions $a_2=0$ and $a_4=0$ determines a critical $E_F$,
with $E_F^c=\alpha(z)T_c$ whenever $T_c \ll W_0$,
such that the transition changes from second order
for $E_F<E_F^c$, to first order in the region $E_F>E_F^c$ where we have $a_4<0$.
We find numerically that
$\cosh \alpha(z)$ varies between 3.4 at $z=2$ and 2 at $z\rightarrow
\infty$.

A typical phase diagram with $z>2$ is shown in
Fig.[\ref{phase_diagram}], where the value of the gap $\Delta_c$ at
the critical temperature, in the region where the transition is
first order is also shown. When the line of first order transitions
is crossed, the electron density jumps discontinuously. For
sufficiently large values of $E_F$ we find that $\Delta_c (E_F )
> E_F$. When the transition line is crossed in this region, the
electron density in the ordered phase is zero. The phase diagram as
function of temperature and electron density is shown in
Fig.[\ref{phase_separation}].

\section{Formation of lattice defects.}
\subsection{Unbinding of dilocations.}

As discussed earlier, ripples, e.g. due pinning to a rough substrate, induce in plane strains. If these strains
are sufficiently large, it will become favorable to relax them by creating lattice dislocations. It is
convenient to view the out of plane deformations as inducing quenched random stresses coupling linearly to the
in plane strain tensor $\tilde u_{ij}$ via an energy density $\sum_{ij} \sigma_{ij} \tilde u_{ij}$. One can then
apply the result of Ref. \onlinecite{CD98b} for the threshold beyond which random stresses generate
dislocations.

Using eq.(\ref{hamil}), the random stress tensor field which renormalizes the fugacity of dislocations is:
\begin{eqnarray}
\sigma_{xx} &= &\frac{\lambda}{2} \left[ \left( \frac{\partial h}{\partial x} \right)^2 + \left( \frac{\partial
h}{\partial y} \right)^2 \right] + \mu \left( \frac{\partial h}{\partial x}
\right)^2 \nonumber \\
\sigma_{yy} &= &\frac{\lambda}{2} \left[ \left( \frac{\partial h}{\partial x} \right)^2 + \left( \frac{\partial
h}{\partial y} \right)^2 \right] + \mu \left( \frac{\partial h}{\partial y}
\right)^2 \nonumber \\
\sigma_{xy} &= &\mu \frac{\partial h}{\partial x} \frac{\partial h}{\partial y}
\end{eqnarray}
We assume that the correlations of this field are given by:
\begin{equation}
\left\langle \sigma_{ij} ( \vec q ) \sigma_{kl} ( - \vec q ) \right\rangle|_{q \to 0} = \left[ \sigma_{\lambda}
\delta_{ij} \delta_{kl} +
 \sigma_{\mu} \left( \delta_{ik} \delta_{lj} + \delta_{il} \delta_{jk}
\right) \right]
\end{equation}
where the parameters $\sigma_{\mu}$ and $\sigma_{\lambda}+ 2 \sigma_\mu$ measure the strength of random shear
stresses and compressional stresses, respectively.

In presence of random stresses an isolated dislocation in a region of size $L$ feels a random potential whose
minima grow typically $\sim - \ln L$. The logarithmic elastic energy cost of creating a dislocation can then be
overcome, and thus dislocations will proliferate at $T=0$, when:
\begin{equation}
\tilde \sigma = \frac{\lambda ( \lambda + 2 \mu ) \sigma_{\mu} + \mu^2 (\sigma_{\lambda} + 2 \sigma_\mu) }{\mu^2
( \lambda +  \mu )^2} \ge \tilde \sigma_c = \frac{a^2}{16 \pi} \label{sigma_dis}
\end{equation}
where we have neglected the effect of screening of the elastic
coefficients by disorder, which have been shown to be small
\footnote{We have also taken into account the factor 2 misprint in
$\sigma$ as defined below Eq. (5) in \cite{CD98b}}). To the same
accuracy this formula holds for all $T < T_m/2$ where $T_m= K_0
a^2/(16 \pi)$ is the KTHNY melting temperature of a pure 2d crystal,
with $K_0=4 \mu (\mu+\lambda)/(\mu+2 \lambda)$, while the threshold
decreases as $\tilde \sigma_c(T)=4 \tilde \sigma_c
\frac{T}{T_m}(1-\frac{T}{T_m})$ at higher $T$. For $\tilde \sigma >
\tilde \sigma_c$ and at $T=0$ the scale $L$ above which dislocation
first appear can be estimated as in \cite{BHPLD} and corresponds to
the total energy cost $\frac{K_0 a^2}{8 \pi} (1 - \sqrt{\tilde
\sigma/\tilde \sigma_c})  \ln (L/l_0) +E_c$ becoming negative. We
have taken into account the dislocation core energy $E_c = E_c^0 +
\frac{K_0 a^2}{8 \pi} \ln (l_0/a)$ at scale $l_0$ ($E_c^0$ denotes
the bare core energy). Because of logarithms this scale can be large
hence it can alternatively be viewed as defining an effective size
dependent threshold $\tilde \sigma_c(L)$. The dislocation density
above this scale can be estimated by arguments similar to
\cite{pldtg}.

The quantities $\sigma_{\lambda}$ and $\sigma_{\mu}$ can be written in terms of the correlations of the function
$f_{ij}$, given in eqs.(\ref{corr}) and (\ref{gauge}):
\begin{eqnarray}
\sigma_{\lambda} &= &\frac{a^2}{\beta^2} \left[ \mu^2 ( f + 2 f' ) + \lambda ( \lambda + 2 \mu ) ( f +
f' ) \right]\nonumber \\
\sigma_{\mu} &= &\frac{a^2}{\beta^2} \mu^2 f' \label{corr2}
\end{eqnarray}
Inserting this result in eq.(\ref{sigma_dis}), and assuming that $\beta , \lambda / \mu \sim O ( 1 )$, we find
that dislocations will proliferate when the height correlations are such that $h^2 / ( l_0 a ) \gtrsim 1$, which
is the same combination of scales which determines the existence of a divergence in the electronic density of
states.

\subsection{Buckling into the third dimension.}
An effect not taken into account above is that dislocations may buckle in the third dimension to lower their
energy. For a free membrane (in the absence of a substrate) this occurs for scales larger than the buckling
radius $R_b$, and below that scale the membrane remains flat and Coulomb gas logarithmic scaling holds. In
principle, for a free membrane in presence of internal in plane random stresses, if $R_b$ is large enough
(values such as $R_b \sim 10^2 \kappa/(K_0 a)$ are quoted in Ref. \cite{nelson_seung}), i.e. if $R_b
> l_0 \gg a$, the above energy estimate setting $L=R_b$ can be used to determine the disorder threshold at which
buckled dislocations would occur. However, if one takes into account the pinning of the height field to the
substrate, the energy calculation of Ref. \cite{nelson_seung}) remains valid for scales smaller than $l_p$, but
must be reexamined for scales larger than $l_p$, a problem left for future study.

\subsection{Gauge fields associated to dislocations.}
Note, finally, that dislocation cores act on the electrons outside
the core as vortices \cite{GGV01} of flux $\Phi = \epsilon \Phi_0 /
3$, where $\Phi_0$ is the quantum unit of flux ($=2 \pi$ in our
units), and $\epsilon = \pm 1$. Hence, the existence of dislocations
will increase the random field due to elastic strains considered so
far. Given a set of dislocations at position $\vec r_n$ and Burgers
charges $\vec b_n$ the resulting effective magnetic field
 can be written $B(\vec
r) = (\Phi_0/3) n(\vec r)$ where $n(\vec r)=\sum_n \epsilon_n
\delta(\vec r - \vec r_n)$ and the signs are given by $\epsilon = 2
\vec b \cdot a_1 \text{mod} 2 \pi$. If positions and signs were
chosen uncorrelated (such as in a quench from infinite temperature)
it would result in a LR correlated random gauge field, i.e $C_B(q)
\sim \Phi_0^2 d^{-2}$ at small $q$ in (\ref{defB}), where $d$ is the
mean distance between defects\footnote{Strictly, $B$ is zero outside
the core and $\vec A$ is a pure (singular) gauge, with correlations
which diverge\cite{K07} as $| \vec{q} |^{-2}$. A coarse grained
field in the whole space can be computed\cite{GGV01}- including the
cores - by assigning minimal flux to each vortex (note, however,
that each of these fluxes can be increased by an integer flux
quantum without changing the eigenenergies while increasing the
degeneracy).}

This procedure however leads to Burgers charge fluctuations growing
as $\pm \sim L$ in an area $L^2$ hence a very large elastic energy,
$L \ln L$. If the system can relax, this energy is screened and the
result is a finite parameter $\sigma$ as defined in
(\ref{estimate}). In cases where the dislocation density is not very
small it can be estimated from a Debye-H\"uckel theory. One
non-equilibrium example is a quench of a pure crystal from a
(moderate) temperature $T_Q>T_m$ to low temperature in which case
$\langle n(\vec q) n(-\vec q) \rangle = T_Q q^2/(E_c^0 q^2 + K_0
a^2)$ hence $\sigma=T_Q \Phi_0^2/(9 \pi K_0 a^2)$. Further
relaxation of $\sim \ln L$ energy would then occur. Another example
is the distribution of dislocations induced by the ripples as in
(\ref{sigma_dis}), with $\tilde \sigma
> \tilde \sigma_c$. Then one estimates \footnote{Due to the relation between flux and Burgers vector the
problem becomes isotropic and one may neglect the vector nature of
the charges} $\langle n(\vec q) n(-\vec q) \rangle = \frac{1}{4} q^2
\tilde \sigma K_0^2 a^2/(E_c^0 q^2 + K_0 a^2)^2$ hence $\sigma=
\tilde \sigma \Phi_0^2/(36 \pi a^2)$. Very near the transition
Debye-H\"uckel does not apply as $\sigma$ vanishes at $\sigma_c$
proportionally to the density of dislocations.

\section{ripples in suspended graphene}

Finally we discuss a possible source for ripples in suspended
graphene\cite{Metal07,Betal08,DSBA08}. Upon etching a preexisting
rough substrate, the rippled graphene sheet would tend to relax to a
flat configuration with higher projected area. This however may be
precluded if the sheet is pinned at its boundaries. Indeed it is
known that fixed connectivity membranes exhibit a buckled state when
constrained at their boundaries by a fixed frame of projected area
$A_f$ smaller than the equilibrium area of the unconstrained
membrane $A$. As discussed in \cite{buckling1} it results in an
additional compressional energy term of the form $\tau \int d^2 \vec
r \sum_i u_{ii}$ and hence implies that the energy of flexural modes
becomes, to lowest order $\frac{1}{2} \int d^2 \vec r [ \kappa
(\nabla^2 h)^2 + \tau (\nabla h)^2]$. In the buckled phase, $A_f<A$,
$\tau <0$, an instability thus develops at scales larger than $\xi_h
\sim (\kappa/|\tau|)^{1/2}$. This phase can be described as a non
homogeneous mixture of pure flat phases with different orientations.
For arbitrary boundary conditions, it is expected to be non trivial
since, contrary to a one dimensional rod, in a polymerized membrane
the in-plane modes cannot fully relax the flexural constraints, i.e.
the transverse part of the flexural strain tensor, $P^T_{i
j}(\nabla) (\partial_i h \partial_j h)$, cannot be relaxed by the
in-plane strain field. While demonstrating the instability is
simple, the full calculation of the resulting shape requires
consideration of non linear terms and is difficult. The problem of
relaxation from a randomly rippled configuration with a fixed frame
constraint deserves further study, in particular the question of
whether there is some memory of the initial ripple pattern.

Note that one may consider, alternatively, unconstrained boundary and apply a tension $- f \int d^2 \vec r
\partial_i u_{i}$. The buckling transition \cite{buckling1} has been mostly studied on the side where the sheet
is stretched (the effective $\tau_R \to 0^+$). It was found that $(A_f-A)/A \sim \tau_R \sim |f|^{1/\delta}
\text{sign}(f)$ and the correlation lengths $\xi_h \sim \xi_u \sim |f|^{-\nu/\delta}$ for flexural and phonon
modes at small $f$. While entropic effects produce non trivial values for these exponents, these may be
observable only at large scales, and at intermediate scales mean field values $\delta=1$, $\nu=1/2$ (discussed
above) are appropriate. It would thus be interesting to study the other side of this transition.

\section{Broadening of Landau levels in presence of a real magnetic field}

 In presence of a real magnetic field $B$ we expect that the
Landau levels will be broadened by the effective field $B_{rip}$
due to the ripples. Alternatively, the local electronic density corresponding to $N$ full Landau
levels is fluctuating according to $n(\rr)=[B+B_{rip}(\rr)] N/\phi_0$,
where $\phi_0$ is the flux quantum. Such density fluctuations were
recently measured \cite{Metal07b} showing $\delta n=\pm 2.3\cdot
10^{11}$cm$^{-2}$ at a field of $11T$ for $N=2,6,10$. In this
section we estimate the contribution of the random gauge field due to ripples to these density
fluctuations.

Consider the density $n$ measured on a length scale $L$, and its probability
distribution, near an average density
$BN/\phi_0$.
We assume first
$l_0>l_B\sqrt{N}$, where $l_B=\sqrt{\phi_0/B}$ is the cyclotron
radius and $l_B\sqrt{N}$ estimates the size of an orbit in the N-th Landau
level. Each Landau orbit has then a random shift $\pm
B_{rip}(\rr)$ where the $\pm$ corresponds to the K and K' valleys
that feel opposite gauge fields. The density distribution is
\begin{equation} P(n)=\int_{r<L} \sum_{\pm}\delta(n-BN/\phi_0\pm
B_{rip}(r)N/\phi_0)d^2r/2L^2 \end{equation}
The average is $\langle n\rangle=BN/\phi_0$ while the variance is:
\begin{equation} \label{variance} \langle \delta
n^2\rangle=\frac{N^2}{\phi_0^2 L^4}\langle
[\int_{r<L}B_{rip}(\rr)d^2r]^2\rangle
=\frac{N^2}{4\pi^2L^4}\langle [\oint A(u)du]^2\rangle
\end{equation}
where $B_{rip}=[\bm{\nabla}\times A]_z\phi_0/2\pi$ and $A$ is the random
gauge field considered in the previous Sections (with the appropriate
change in units) and the contour encloses the area of measurement.
To estimate the variance in (\ref{variance}) we use
Eq. (\ref{estimate}) with a cutoff $\exp{(-q^2l_0^2/2)}$.
In real space it corresponds to  $\langle
A_i(\rr)A_j(\rr)\rangle \sim (\sigma/2l_0^2)\exp{[-(\rr-\rr')^2/2l_0^2]}$
where we neglect \footnote{a similar estimate can
be done using the magnetic field and yields the same result}
the transversality constraint on $A$. It yields:
\begin{equation}\label{dn}
\langle \delta n^2\rangle_L \approx \frac{N^2\sigma}{4\pi l_0 L^3}
\end{equation}

The experimentally more relevant case is $l_0<l_B\sqrt{N}$. In
this case we argue that ${\bf A}(\rr)$ can be replaced by its
average within a Landau state, so that an average ripple field is
\begin{equation}
B_{av}(\rr)=\frac{1}{2\pi Nl_B^2}\int d^2r_0
B_{rip}(\rr_0)\eexp{-(\rr-\rr_0)^2/2Nl_B^2}
\end{equation}
and its Fourier transform is
$B_{av}(q)=B_{rip}(q)\exp(-q^2Nl_B^2/2)$. This replaces
$l_0\rightarrow l_B\sqrt{N}$ in Eq. (\ref{dn}), and identifying
$L$ with the tip size $l_{tip}$ in the experiment \cite{Metal07b},
we obtain,
\begin{equation}\label{dn1}
\langle \delta n^2\rangle_L \approx N^{3/2}\frac{\sigma}{4\pi l_B
l_{tip}^3}
\end{equation}
Using \cite{Metal07b} $l_0\approx 100$nm  and $l_B\approx 10$nm,
Eq. (\ref{dn1}) yields numbers consistent with the experiment,
except for the $N$ dependence. Note that an even weaker dependence
in $N$ ($\delta n \sim N^{1/4}$) is obtained if one assumes that
$L=l_B \sqrt{N}$ is the only averaging scale.

It is also interesting to estimate the energy broadening $\delta
\epsilon_N$ of the Landau levels, which in the absence of ripples
have energies $\epsilon_N = v_F\sqrt{2e} \sqrt{B N'}$ with
$N=4N'+2$. The field associated with the ripples changes locally the
energy of the Landau levels, which become $\epsilon_N = v_F\sqrt{2e}
\sqrt{( B \pm \delta B_{rip}) N'}$, where $\delta B_{rip}$ is the
average value of $B_{rip}$ in the region occupied by the Landau
level. A similar calculation then yields the estimate for $N>2$
\begin{equation}
\delta \epsilon_{N'}\approx v_F\left(\frac{\sigma}{32\pi l_0
l_B}\right)^{1/2}N^{-1/4}
\end{equation}
in the regime $l_B\sqrt{N}>l_0$.

Finally we note, that the $N'=0$ level has no broadening at all.
This remarkable result is obtained by factorizing the $N'=0$
eigenstates of the free Dirac system in a magnetic field with the
well known zero energy solutions of the random gauge problem
\cite{LFSG94,HD02}. This set has the proper Landau degeneracy and
is therefore an exact solution for the zero energy Landau level
with random gauge.

\section{Conclusions.}
We have analyzed the effect of random gauge fields on the electronic
structure of corrugated graphene. We find that the local density of states
diverges at the Dirac energy $E=0$, as $\rho ( E ) \propto E^{2/z-1}$, with
$z>2$, for sufficiently strong disorder. The scale of height fluctuations,
$h$ should satisfy $\beta h^2 / ( l a ) \gtrsim 1$, where $\beta \sim 1-2 $ gives the
coupling between the electrons and the lattice strains, $l$ is the typical
spatial scale of the disorder, and $a$ is the lattice constant.

A divergence in the density of non interacting density of states
implies the existence of instabilities in the presence of
electron-electron interactions. We have analyzed the possibility
that a gap will open at low temperatures, depleting the low energy
density of states. We have found a first order transition to an
ordered state at large $E_F$. This discontinuous transition, in
turn, implies electronic phase separation.

When the strains which induce the gauge potential are sufficiently
strong, they can lead to an instability, and the formation of
lattice dislocations. This change takes place for $C ( \lambda ,
\mu ) h^2 / ( l a ) \gtrsim 1$, where $C ( \lambda , \mu ) \sim 1$
is a dimensionless parameter which depends on the elastic
constants of the material.

We have described the main features of the buckling instability
which may arise in suspended systems under compression. Finally, we
analyze the changes induced in the Landau levels induced by a
magnetic field by the gauge potential associated to ripples and show
correspondence with experimental data \cite{Metal07b.}

Our analysis is consistent with previous work on the changes in the
electronic density of states in graphene in the presence of
ripples\cite{GKV07} (see also\cite{WBKL07}). A transition to a state
magnetically ordered in highly disordered systems agrees with the observation
of magnetism in irradiated graphite samples\cite{Eetal03}. The existence
of charge inhomogeneities, due to electronic phase separation, can help to
explain the observation of charge puddles when the Fermi energy is close to
the Dirac energy\cite{Metal07b}.
\section{Acknowledgments.} This work was supported by MEC (Spain) through grant
FIS2005-05478-C02-01, the Comunidad de Madrid, through the program
CITECNOMIK, CM2006-S-0505-ESP-0337, the European Union Contract
12881 (NEST), ANR program 05-BLAN-0099-01 and the DIP German Israeli
program. B.H. and F. G. thank the Ecole Normale Sup\'{e}rieure for
hospitality and for support during part of this work.


\begin{thebibliography}{37}

\expandafter\ifx\csname natexlab\endcsname\relax\def\natexlab#1{#1}\fi
\expandafter\ifx\csname bibnamefont\endcsname\relax
  \def\bibnamefont#1{#1}\fi
\expandafter\ifx\csname bibfnamefont\endcsname\relax
  \def\bibfnamefont#1{#1}\fi
\expandafter\ifx\csname citenamefont\endcsname\relax
  \def\citenamefont#1{#1}\fi
\expandafter\ifx\csname url\endcsname\relax
  \def\url#1{\texttt{#1}}\fi
\expandafter\ifx\csname urlprefix\endcsname\relax\def\urlprefix{URL }\fi
\providecommand{\bibinfo}[2]{#2}
\providecommand{\eprint}[2][]{\url{#2}}

\bibitem[{\citenamefont{Novoselov et~al.}(2004)\citenamefont{Novoselov, Geim,
  Morozov, Jiang, Zhang, Dubonos, Gregorieva, and Firsov}}]{Netal04}
\bibinfo{author}{\bibfnamefont{K.~S.} \bibnamefont{Novoselov}},
  \bibinfo{author}{\bibfnamefont{A.~K.} \bibnamefont{Geim}},
  \bibinfo{author}{\bibfnamefont{S.~V.} \bibnamefont{Morozov}},
  \bibinfo{author}{\bibfnamefont{D.}~\bibnamefont{Jiang}},
  \bibinfo{author}{\bibfnamefont{Y.}~\bibnamefont{Zhang}},
  \bibinfo{author}{\bibfnamefont{S.~V.} \bibnamefont{Dubonos}},
  \bibinfo{author}{\bibfnamefont{I.~V.} \bibnamefont{Gregorieva}},
  \bibnamefont{and} \bibinfo{author}{\bibfnamefont{A.~A.}
  \bibnamefont{Firsov}}, \bibinfo{journal}{Science}
  \textbf{\bibinfo{volume}{306}}, \bibinfo{pages}{666} (\bibinfo{year}{2004}).

\bibitem[{\citenamefont{Novoselov et~al.}(2005)\citenamefont{Novoselov, Jiang,
  Schedin, Booth, Khotkevich, Morozov, and Geim}}]{Netal05b}
\bibinfo{author}{\bibfnamefont{K.~S.} \bibnamefont{Novoselov}},
  \bibinfo{author}{\bibfnamefont{D.}~\bibnamefont{Jiang}},
  \bibinfo{author}{\bibfnamefont{F.}~\bibnamefont{Schedin}},
  \bibinfo{author}{\bibfnamefont{T.~J.} \bibnamefont{Booth}},
  \bibinfo{author}{\bibfnamefont{V.~V.} \bibnamefont{Khotkevich}},
  \bibinfo{author}{\bibfnamefont{S.~V.} \bibnamefont{Morozov}},
  \bibnamefont{and} \bibinfo{author}{\bibfnamefont{A.~K.} \bibnamefont{Geim}},
  \bibinfo{journal}{Proc. Nat. Acad. Sc.} \textbf{\bibinfo{volume}{102}},
  \bibinfo{pages}{10451} (\bibinfo{year}{2005}).

\bibitem[{\citenamefont{Geim and Novoselov}(2007)}]{GN07}
\bibinfo{author}{\bibfnamefont{A.~K.} \bibnamefont{Geim}} \bibnamefont{and}
  \bibinfo{author}{\bibfnamefont{K.~S.} \bibnamefont{Novoselov}},
  \bibinfo{journal}{Nature Materials} \textbf{\bibinfo{volume}{6}},
  \bibinfo{pages}{183} (\bibinfo{year}{2007}).

\bibitem[{\citenamefont{{Castro Neto} et~al.}(2007)\citenamefont{{Castro Neto},
  Guinea, Peres, Novoselov, and Geim}}]{NGPNG07}
\bibinfo{author}{\bibfnamefont{A.~H.} \bibnamefont{{Castro Neto}}},
  \bibinfo{author}{\bibfnamefont{F.}~\bibnamefont{Guinea}},
  \bibinfo{author}{\bibfnamefont{N.~M.~R.} \bibnamefont{Peres}},
  \bibinfo{author}{\bibfnamefont{K.~S.} \bibnamefont{Novoselov}},
  \bibnamefont{and} \bibinfo{author}{\bibfnamefont{A.~K.} \bibnamefont{Geim}}
  (\bibinfo{year}{2007}), \eprint{arXiv:0709.1163}.

\bibitem[{\citenamefont{Gonz\'alez et~al.}(1992)\citenamefont{Gonz\'alez,
  Guinea, and Vozmediano}}]{GGV92}
\bibinfo{author}{\bibfnamefont{J.}~\bibnamefont{Gonz\'alez}},
  \bibinfo{author}{\bibfnamefont{F.}~\bibnamefont{Guinea}}, \bibnamefont{and}
  \bibinfo{author}{\bibfnamefont{M.~A.~H.} \bibnamefont{Vozmediano}},
  \bibinfo{journal}{Phys. Rev. Lett.} \textbf{\bibinfo{volume}{69}},
  \bibinfo{pages}{172} (\bibinfo{year}{1992}).

\bibitem[{\citenamefont{Gonz\'alez et~al.}(1993)\citenamefont{Gonz\'alez,
  Guinea, and Vozmediano}}]{GGV93b}
\bibinfo{author}{\bibfnamefont{J.}~\bibnamefont{Gonz\'alez}},
  \bibinfo{author}{\bibfnamefont{F.}~\bibnamefont{Guinea}}, \bibnamefont{and}
  \bibinfo{author}{\bibfnamefont{M.~A.~H.} \bibnamefont{Vozmediano}},
  \bibinfo{journal}{Nucl. Phys. B} \textbf{\bibinfo{volume}{406 [FS]}},
  \bibinfo{pages}{771} (\bibinfo{year}{1993}).

\bibitem[{\citenamefont{Morozov et~al.}(2006)\citenamefont{Morozov, Novoselov,
  Katsnelson, Schedin, Ponomarenko, Jiang, and Geim}}]{Metal06}
\bibinfo{author}{\bibfnamefont{S.~V.} \bibnamefont{Morozov}},
  \bibinfo{author}{\bibfnamefont{K.~S.} \bibnamefont{Novoselov}},
  \bibinfo{author}{\bibfnamefont{M.~I.} \bibnamefont{Katsnelson}},
  \bibinfo{author}{\bibfnamefont{F.}~\bibnamefont{Schedin}},
  \bibinfo{author}{\bibfnamefont{L.~A.} \bibnamefont{Ponomarenko}},
  \bibinfo{author}{\bibfnamefont{D.}~\bibnamefont{Jiang}}, \bibnamefont{and}
  \bibinfo{author}{\bibfnamefont{A.~K.} \bibnamefont{Geim}},
  \bibinfo{journal}{Phys. Rev. Lett.} \textbf{\bibinfo{volume}{97}},
  \bibinfo{pages}{016801} (\bibinfo{year}{2006}).

\bibitem[{\citenamefont{Morpurgo and Guinea}(2006)}]{MG06}
\bibinfo{author}{\bibfnamefont{A.}~\bibnamefont{Morpurgo}} \bibnamefont{and}
  \bibinfo{author}{\bibfnamefont{F.}~\bibnamefont{Guinea}},
  \bibinfo{journal}{Phys. Rev. Lett.} \textbf{\bibinfo{volume}{97}},
  \bibinfo{pages}{196804} (\bibinfo{year}{2006}).

\bibitem[{\citenamefont{{Castro Neto} and Kim}(2007)}]{NK07}
\bibinfo{author}{\bibfnamefont{A.~H.} \bibnamefont{{Castro Neto}}}
  \bibnamefont{and} \bibinfo{author}{\bibfnamefont{E.-A.} \bibnamefont{Kim}}
  (\bibinfo{year}{2007}), \eprint{arXiv:cond-mat/0702562}.

\bibitem[{\citenamefont{Meyer et~al.}(2007)\citenamefont{Meyer, Geim,
  Katsnelson, Novoselov, Booth, and Roth}}]{Metal07}
\bibinfo{author}{\bibfnamefont{J.~C.} \bibnamefont{Meyer}},
  \bibinfo{author}{\bibfnamefont{A.~K.} \bibnamefont{Geim}},
  \bibinfo{author}{\bibfnamefont{M.~I.} \bibnamefont{Katsnelson}},
  \bibinfo{author}{\bibfnamefont{K.~S.} \bibnamefont{Novoselov}},
  \bibinfo{author}{\bibfnamefont{T.~J.} \bibnamefont{Booth}}, \bibnamefont{and}
  \bibinfo{author}{\bibfnamefont{S.}~\bibnamefont{Roth}},
  \bibinfo{journal}{Nature} \textbf{\bibinfo{volume}{446}}, \bibinfo{pages}{60}
  (\bibinfo{year}{2007}).



\bibitem[{\citenamefont{Sabio et~al.}(2007)\citenamefont{Sabio, Seo\'anez,
  Fratini, Guinea, {Castro Neto}, and Sols}}]{Setal07b}
\bibinfo{author}{\bibfnamefont{J.}~\bibnamefont{Sabio}},
  \bibinfo{author}{\bibfnamefont{C.}~\bibnamefont{Seo\'anez}},
  \bibinfo{author}{\bibfnamefont{S.}~\bibnamefont{Fratini}},
  \bibinfo{author}{\bibfnamefont{F.}~\bibnamefont{Guinea}},
  \bibinfo{author}{\bibfnamefont{A.~H.} \bibnamefont{{Castro Neto}}},
  \bibnamefont{and} \bibinfo{author}{\bibfnamefont{F.}~\bibnamefont{Sols}}
  (\bibinfo{year}{2007}), \eprint{arXiv:0712.2232}.

\bibitem[{\citenamefont{Cho and Fuhrer}(2007)}]{CF07}
\bibinfo{author}{\bibfnamefont{S.}~\bibnamefont{Cho}} \bibnamefont{and}
  \bibinfo{author}{\bibfnamefont{M.~S.} \bibnamefont{Fuhrer}}
  (\bibinfo{year}{2007}), \eprint{arXiv:0705.3239}.

\bibitem{Vetal08}
A. L. V\'azquez de Parga, F. Calleja, B. Borca, M. C. Passeggi, J.
J. Hinarejos, F. Guinea, and R. Miranda, Phys. Rev. Lett. {\bf 100},
056807 (2008).

\bibitem[{\citenamefont{Ludwig et~al.}(1994)\citenamefont{Ludwig, Fisher,
  Shankar, and Grinstein}}]{LFSG94}
\bibinfo{author}{\bibfnamefont{A.~W.} \bibnamefont{Ludwig}},
  \bibinfo{author}{\bibfnamefont{M.~P.~A.} \bibnamefont{Fisher}},
  \bibinfo{author}{\bibfnamefont{R.}~\bibnamefont{Shankar}}, \bibnamefont{and}
  \bibinfo{author}{\bibfnamefont{G.}~\bibnamefont{Grinstein}},
  \bibinfo{journal}{Phys. Rev. B} \textbf{\bibinfo{volume}{50}},
  \bibinfo{pages}{7526} (\bibinfo{year}{1994}).



\bibitem[{\citenamefont{Chamon et~al.}(1996)\citenamefont{Chamon, Mudry, and
  Wen}}]{CMW96}
\bibinfo{author}{\bibfnamefont{C.}~\bibnamefont{Chamon}},
  \bibinfo{author}{\bibfnamefont{C.}~\bibnamefont{Mudry}}, \bibnamefont{and}
  \bibinfo{author}{\bibfnamefont{X.-G.} \bibnamefont{Wen}},
  \bibinfo{journal}{Phys. Rev. Lett.} \textbf{\bibinfo{volume}{77}},
  \bibinfo{pages}{4194} (\bibinfo{year}{1996}).

\bibitem[{\citenamefont{Castillo et~al.}(1997)\citenamefont{Castillo, {de C.
  Chamon}, Fradkin, Goldbart, and Mudry}}]{CCFGM97}
\bibinfo{author}{\bibfnamefont{H.~E.} \bibnamefont{Castillo}},
  \bibinfo{author}{\bibfnamefont{C.}~\bibnamefont{{de C. Chamon}}},
  \bibinfo{author}{\bibfnamefont{E.}~\bibnamefont{Fradkin}},
  \bibinfo{author}{\bibfnamefont{P.~M.} \bibnamefont{Goldbart}},
  \bibnamefont{and} \bibinfo{author}{\bibfnamefont{C.}~\bibnamefont{Mudry}},
  \bibinfo{journal}{Phys. Rev. B} \textbf{\bibinfo{volume}{56}},
  \bibinfo{pages}{10668} (\bibinfo{year}{1997}).

\bibitem[{\citenamefont{Horovitz and {Le Doussal}}(2002)}]{HD02}
\bibinfo{author}{\bibfnamefont{B.}~\bibnamefont{Horovitz}} \bibnamefont{and}
  \bibinfo{author}{\bibfnamefont{P.}~\bibnamefont{{Le Doussal}}},
  \bibinfo{journal}{Phys. Rev. B} \textbf{\bibinfo{volume}{65}},
  \bibinfo{pages}{125323} (\bibinfo{year}{2002}).

\bibitem[{\citenamefont{Khveshchenko}(2007)}]{K07}
\bibinfo{author}{\bibfnamefont{D.~V.} \bibnamefont{Khveshchenko}},
  \bibinfo{journal}{Phys. Rev. B} \textbf{\bibinfo{volume}{75}},
  \bibinfo{pages}{153405} (\bibinfo{year}{2007}).

\bibitem[{\citenamefont{Carpentier and {Le Doussal}}(2000)}]{CD00}
\bibinfo{author}{\bibfnamefont{D.}~\bibnamefont{Carpentier}} \bibnamefont{and}
  \bibinfo{author}{\bibfnamefont{P.}~\bibnamefont{{Le Doussal}}},
  \bibinfo{journal}{Nucl. Phys. B} \textbf{\bibinfo{volume}{588}},
  \bibinfo{pages}{565} (\bibinfo{year}{2000}).

\bibitem[{\citenamefont{We}(1999)}]{W99}
\bibinfo{author}{\bibfnamefont{J.}~\bibnamefont{We}}, \bibinfo{journal}{Phys.
  Rev. B} \textbf{\bibinfo{volume}{60}}, \bibinfo{pages}{8290}
  (\bibinfo{year}{1999}).

\bibitem[{\citenamefont{Stauber et~al.}(2005)\citenamefont{Stauber, Guinea, and
  Vozmediano}}]{SGV05}
\bibinfo{author}{\bibfnamefont{T.}~\bibnamefont{Stauber}},
  \bibinfo{author}{\bibfnamefont{F.}~\bibnamefont{Guinea}}, \bibnamefont{and}
  \bibinfo{author}{\bibfnamefont{M.~A.~H.} \bibnamefont{Vozmediano}},
  \bibinfo{journal}{Phys. Rev. B} \textbf{\bibinfo{volume}{71}},
  \bibinfo{pages}{041406} (\bibinfo{year}{2005}).

\bibitem{HJV08}
I. F. Herbut, V. Juricic, and O. Vafek, Phys. Rev. Lett. {\bf 100},
046403 (2008).

\bibitem[{\citenamefont{Carpentier and {Le Doussal}}(1998)}]{CD98b}
\bibinfo{author}{\bibfnamefont{D.}~\bibnamefont{Carpentier}} \bibnamefont{and}
  \bibinfo{author}{\bibfnamefont{P.}~\bibnamefont{{Le Doussal}}},
  \bibinfo{journal}{Phys. Rev. Lett.} \textbf{\bibinfo{volume}{81}},
  \bibinfo{pages}{1881} (\bibinfo{year}{1998}).

\bibitem[{\citenamefont{Martin et~al.}(2007)\citenamefont{Martin, Akerman,
  Ulbricht, Lohmann, Smet, von Klitzing, and Yacoby}}]{Metal07b}
\bibinfo{author}{\bibfnamefont{J.}~\bibnamefont{Martin}},
  \bibinfo{author}{\bibfnamefont{N.}~\bibnamefont{Akerman}},
  \bibinfo{author}{\bibfnamefont{G.}~\bibnamefont{Ulbricht}},
  \bibinfo{author}{\bibfnamefont{T.}~\bibnamefont{Lohmann}},
  \bibinfo{author}{\bibfnamefont{J.~H.} \bibnamefont{Smet}},
  \bibinfo{author}{\bibfnamefont{K.}~\bibnamefont{von Klitzing}},
  \bibnamefont{and} \bibinfo{author}{\bibfnamefont{A.}~\bibnamefont{Yacoby}},
  \bibinfo{journal}{Nature Physics} {\bf 4}, 144
  (\bibinfo{year}{2008}).

\bibitem[{\citenamefont{Suzuura and Ando}(2002)}]{SA02b}
\bibinfo{author}{\bibfnamefont{H.}~\bibnamefont{Suzuura}} \bibnamefont{and}
  \bibinfo{author}{\bibfnamefont{T.}~\bibnamefont{Ando}},
  \bibinfo{journal}{Phys. Rev. B} \textbf{\bibinfo{volume}{65}},
  \bibinfo{pages}{235412} (\bibinfo{year}{2002}).

\bibitem[{\citenamefont{{Ma\~nes}}(2007)}]{M07}
\bibinfo{author}{\bibfnamefont{J.~L.} \bibnamefont{{Ma\~nes}}},
  \bibinfo{journal}{Phys. Rev. B} \textbf{\bibinfo{volume}{76}},
  \bibinfo{pages}{045430} (\bibinfo{year}{2007}).

\bibitem[{\citenamefont{Guinea}(1981)}]{G81}
\bibinfo{author}{\bibfnamefont{F.}~\bibnamefont{Guinea}}, \bibinfo{journal}{J.
  Phys. C: Condens. Matt.} \textbf{\bibinfo{volume}{14}}, \bibinfo{pages}{3345}
  (\bibinfo{year}{1981}).

\bibitem[{\citenamefont{Heeger et~al.}(1988)\citenamefont{Heeger, Kivelson,
  Schrieffer, and Su}}]{HKSS88}
\bibinfo{author}{\bibfnamefont{A.~J.} \bibnamefont{Heeger}},
  \bibinfo{author}{\bibfnamefont{S.}~\bibnamefont{Kivelson}},
  \bibinfo{author}{\bibfnamefont{J.~R.} \bibnamefont{Schrieffer}},
  \bibnamefont{and} \bibinfo{author}{\bibfnamefont{W.~P.} \bibnamefont{Su}},
  \bibinfo{journal}{Rev. Mod. Phys.} \textbf{\bibinfo{volume}{60}},
  \bibinfo{pages}{781} (\bibinfo{year}{1988}).

\bibitem[{\citenamefont{Stolyarova et~al.}(2007)\citenamefont{Stolyarova, Rim,
  Ryu, Maultzsch, Kim, Brus, Heinz, Hybertsen, and Flynn}}]{Setal07}
\bibinfo{author}{\bibfnamefont{E.}~\bibnamefont{Stolyarova}},
  \bibinfo{author}{\bibfnamefont{K.~T.} \bibnamefont{Rim}},
  \bibinfo{author}{\bibfnamefont{S.}~\bibnamefont{Ryu}},
  \bibinfo{author}{\bibfnamefont{J.}~\bibnamefont{Maultzsch}},
  \bibinfo{author}{\bibfnamefont{P.}~\bibnamefont{Kim}},
  \bibinfo{author}{\bibfnamefont{L.~E.} \bibnamefont{Brus}},
  \bibinfo{author}{\bibfnamefont{T.~F.} \bibnamefont{Heinz}},
  \bibinfo{author}{\bibfnamefont{M.~S.} \bibnamefont{Hybertsen}},
  \bibnamefont{and} \bibinfo{author}{\bibfnamefont{G.~W.} \bibnamefont{Flynn}},
  \bibinfo{journal}{Proc. Natl. Acad. Sci. USA} \textbf{\bibinfo{volume}{104}},
  \bibinfo{pages}{9209} (\bibinfo{year}{2007}).

\bibitem[{\citenamefont{Fuchs and Lederer}(2007)}]{FL07}
\bibinfo{author}{\bibfnamefont{J.-N.} \bibnamefont{Fuchs}} \bibnamefont{and}
  \bibinfo{author}{\bibfnamefont{P.}~\bibnamefont{Lederer}},
  \bibinfo{journal}{Phys. Rev. Lett.} \textbf{\bibinfo{volume}{98}},
  \bibinfo{pages}{016803} (\bibinfo{year}{2007}).

\bibitem[{\citenamefont{Sorella and Tosatti}(1992)}]{ST92}
\bibinfo{author}{\bibfnamefont{S.}~\bibnamefont{Sorella}} \bibnamefont{and}
  \bibinfo{author}{\bibfnamefont{E.}~\bibnamefont{Tosatti}},
  \bibinfo{journal}{Europhys. Lett.} \textbf{\bibinfo{volume}{19}},
  \bibinfo{pages}{699} (\bibinfo{year}{1992}).

\bibitem[{\citenamefont{Guinea et~al.}(2001)\citenamefont{Guinea, Gonz\'alez,
  and Vozmediano}}]{GGV01}
\bibinfo{author}{\bibfnamefont{F.}~\bibnamefont{Guinea}},
  \bibinfo{author}{\bibfnamefont{J.}~\bibnamefont{Gonz\'alez}},
  \bibnamefont{and} \bibinfo{author}{\bibfnamefont{M.~A.~H.}
  \bibnamefont{Vozmediano}}, \bibinfo{journal}{Phys. Rev. B}
  \textbf{\bibinfo{volume}{63}}, \bibinfo{pages}{134421}
  (\bibinfo{year}{2001}).

\bibitem[{\citenamefont{Peres et~al.}(2004)\citenamefont{Peres, Ara\'ujo, and
  Bozi}}]{PAB04}
\bibinfo{author}{\bibfnamefont{N.~M.~R.} \bibnamefont{Peres}},
  \bibinfo{author}{\bibfnamefont{M.~A.~N.} \bibnamefont{Ara\'ujo}},
  \bibnamefont{and} \bibinfo{author}{\bibfnamefont{D.}~\bibnamefont{Bozi}},
  \bibinfo{journal}{Phys. Rev. B} \textbf{\bibinfo{volume}{70}},
  \bibinfo{pages}{195122} (\bibinfo{year}{2004}).

\bibitem[{\citenamefont{Herbut}(2006)}]{H06}
\bibinfo{author}{\bibfnamefont{I.}~\bibnamefont{Herbut}},
  \bibinfo{journal}{Phys. Rev. Lett.} \textbf{\bibinfo{volume}{97}},
  \bibinfo{pages}{146401} (\bibinfo{year}{2006}).

\bibitem[{\citenamefont{Guinea et~al.}(2000)\citenamefont{Guinea,
  G\'omez-Santos, and Arovas}}]{GGA00}
\bibinfo{author}{\bibfnamefont{F.}~\bibnamefont{Guinea}},
  \bibinfo{author}{\bibfnamefont{G.}~\bibnamefont{G\'omez-Santos}},
  \bibnamefont{and} \bibinfo{author}{\bibfnamefont{D.~P.}
  \bibnamefont{Arovas}}, \bibinfo{journal}{Phys. Rev. B}
  \textbf{\bibinfo{volume}{62}}, \bibinfo{pages}{391} (\bibinfo{year}{2000}).

\bibitem{GKV07}
F. Guinea, M. I. Katsnelson, and M. A. H. Vozmediano, Phys. Rev. B
{\bf 77}, 075422 (2008).

\bibitem[{\citenamefont{Wehling et~al.}(2007)\citenamefont{Wehling, Balatsky,
  Katsnelson, and Lichtenstein}}]{WBKL07}
\bibinfo{author}{\bibfnamefont{T.~O.} \bibnamefont{Wehling}},
  \bibinfo{author}{\bibfnamefont{A.~V.} \bibnamefont{Balatsky}},
  \bibinfo{author}{\bibfnamefont{M.~I.} \bibnamefont{Katsnelson}},
  \bibnamefont{and} \bibinfo{author}{\bibfnamefont{A.~I.}
  \bibnamefont{Lichtenstein}} (\bibinfo{year}{2007}), \eprint{arXiv:0710.5828}.

\bibitem[{\citenamefont{Esquinazi et~al.}(2003)\citenamefont{Esquinazi,
  Spemann, H\"ohne, Setzer, Han, and Butz}}]{Eetal03}
\bibinfo{author}{\bibfnamefont{P.}~\bibnamefont{Esquinazi}},
  \bibinfo{author}{\bibfnamefont{D.}~\bibnamefont{Spemann}},
  \bibinfo{author}{\bibfnamefont{R.}~\bibnamefont{H\"ohne}},
  \bibinfo{author}{\bibfnamefont{A.}~\bibnamefont{Setzer}},
  \bibinfo{author}{\bibfnamefont{K.-H.} \bibnamefont{Han}}, \bibnamefont{and}
  \bibinfo{author}{\bibfnamefont{T.}~\bibnamefont{Butz}},
  \bibinfo{journal}{Phys. Rev. Lett.} \textbf{\bibinfo{volume}{91}},
  \bibinfo{pages}{227201} (\bibinfo{year}{2003}).

\bibitem{radz}
P. Le Doussal and L. Radzihovsky, Phys. Rev. Letters {\bf 69} 1209 (1992).


\bibitem{nelson_seung}
H. S. Seung and D. R. Nelson, Phys. Rev. A {\bf 38} 1005 1988).

\bibitem{BHPLD}
B. Horovitz and P. Le Doussal, condmat/0410019. Phys. Rev. B 71,
134202 (2005).

\bibitem{pldtg}
P. Le Doussal and T. Giamarchi, cond-mat/9810218, Physica C {\bf
331} 233 (2000)

\bibitem{Betal08}
 K. I. Bolotin, K. J. Sikes, Z. Jiang, G. Fudenberg, J. Hone, P. Kim, and H. L.
 Stormer, arXiv:0802.2389.

\bibitem{DSBA08}
X. Du, I. Skachko, A. Barker, and E. Y. Andrei, arXiv:0802.2933.

\bibitem{buckling1}
E. Guitter et al., Phys. Rev. Lett. {\bf 61} 2949 (1988) and J.
Phys. France {\bf 50} 1787 (1989). J. A. Aronovitz, L. Golubovic and
T.C. Lubensky, J. Phys. France {\bf 50} 609 (1989).



\end{thebibliography}

\end{document}